\documentclass[a4paper,twocolumn,11pt,unpublished
]{quantumarticle}
\pdfoutput=1
\usepackage[utf8]{inputenc}
\usepackage[english]{babel}
\usepackage[T1]{fontenc}
\usepackage{amsmath}
\usepackage{hyperref}

\usepackage{tikz}
\usepackage{lipsum}

\usepackage{braket}
\usepackage{mathtools}
\usepackage[numbers,sort&compress]{natbib}

\begin{document}

\title{Quantum illumination with asymmetrically squeezed two-mode light}

\author{Yonggi Jo}\email{yonggi@add.re.kr}\orcid{0000-0002-2992-304X}
\author{Taek Jeong}
\author{Junghyun Kim}
\author{Duk Y. Kim}
\author{Yong Sup Ihn}\orcid{0000-0003-3883-4062}
\author{Zaeill Kim}
\author{Su-Yong Lee}\email[]{suyong2@add.re.kr}
\affiliation{Agency for Defense Development, Daejeon 34186, South Korea}
\maketitle

\begin{abstract}
We propose Gaussian quantum illumination(QI) protocol exploiting asymmetrically squeezed two-mode(ASTM) state that is generated by applying single-mode squeezing operations on each mode of an initial two-mode squeezed vacuum(TMSV) state, in order to overcome the limited brightness of a TMSV state. We show that the performance of the optimal receiver is enhanced by local squeezing operation on a signal mode whereas the performance of a realistic receiver can be enhanced by local squeezing operations on both input modes. Under a fixed mean photon number of the signal mode, the ASTM state can be close to the TMSV state in the performance of QI while there is a threshold of beating classical illumination in the mean photon number of the initial TMSV state. We also verify that quantum discord cannot be a resource of quantum advantage in the Gaussian QI using the ASTM state, which is a counterexample of a previous claim.
\end{abstract}

\section{Introduction}

Quantum illumination(QI) is a novel remote target detection scheme \cite{Lloyd2008,Tan2008} which exploits quantum correlation under a very noisy channel. QI presents a better target detection performance compared to its classical counterpart, called classical illumination(CI), with the same transmission energy. After the first proposal using single-photon level states \cite{Lloyd2008}, there have been many studies in QI \cite{Tan2008,Luong2020-1,Guha2009,SL09, Wilde2017, Lopaeva2013, Zhang2015, England2019, Luong2019,Barzanjeh2015, Barzanjeh2020, Karsa2020, Devi, Ragy, Zhang14, Sanz, Liu17, Weedbrook, Bradshaw2017, Zubairy, Stefano19, Palma, Ray, Sun, Ranjith, Sandbo, Aguilar, Sussman, Lee, Zhuang2017, Karsa2020-1,Yung,Shapiro2019,Guha2009-1,Jo2021,Lee2021,Daum2020,Lanzagorta2020,Bourassa2020,Bourassa2020-1,Blakely2021,Cai2020,Zhuang2017-1,Bradshaw2021}. Among them, it is realistic to use Gaussian states as input resources so that it is called Gaussian QI \cite{Tan2008,SL09,Barzanjeh2015,Wilde2017,Karsa2020}, where it is typical to use a two-mode squeezed vacuum(TMSV) state consisting of signal and idler modes. In QI, the signal mode is sent to a target for detection, and the idler mode is retained ideally. After the return mode from the target is jointly measured with the idler mode, the decision is made in terms of the measurement outcomes. There were some proposals of QI receivers \cite{Guha2009,Guha2009-1,Zhuang2017,Karsa2020-1,Jo2021,Lee2021} which demonstrate the quantum advantage of Gaussian QI over CI. Experimentally, QI was implemented in laboratories not only using optical frequencies \cite{Lopaeva2013,Zhang2015,England2019} but also using microwave ones \cite{Luong2019,Barzanjeh2020}.

Intuitively, increasing the signal energy enhances the performance of a target detection protocol. The energy of a TMSV state increases with a squeezing parameter, such that highly squeezed light would provide better performance in target detection. Since the reported mean photon number of a single-mode squeezed vacuum state is roughly up to $7.41$ in an optical frequency \cite{Vahlbruch2016} and $1.19$ in a microwave one \cite{Dassonneville2021}, it is a demanding technology of generating a bright TMSV state, i.e., a TMSV state with a large squeezing parameter. To overcome the issue, there was a study on signal amplification of a TMSV state \cite{Bourassa2020}, where the two-mode correlation of an equally amplified state was investigated for the performance of a quantum-enhanced noise radar which measures the signal and idler modes separately.

In this paper, we propose a QI protocol with asymmetrically squeezed two-mode(ASTM) light that is produced by independently performing single-mode squeezing operations on each mode of a TMSV state. The local squeezing operation increases the mean photon number on each mode of the TMSV state, which is feasible rather than attaining a bright TMSV state. The performance of our QI is investigated under the quantum Chernoff bound(QCB) \cite{Audenaert2007,Calsamiglia2008} and the receiver operating characteristic(ROC), where the former assumes the optimal QI receiver and the latter does a realistic one. We show that the additional local squeezing can improve the performance of QI, resulting in quantum advantage compared to CI with the same signal energy in the both receivers. Additionally, our QI provides a counterexample against the claim that the quantum discord is a resource of the quantum advantage in QI \cite{Weedbrook,Bradshaw2017}. Previously it was shown that the quantum advantage in a single-shot target detection with discrete variables cannot be characterized by the quantum discord solely \cite{Yung}. Here, we report an example that the quantum advantage in Gaussian QI cannot be characterized with quantum discord.

This paper is organized as follows. In Sec.~\ref{SecASTMQI} we introduce a basic concept of QI with an ASTM state. In Sec.~\ref{SecPA} we compare the performance of the conventional Gaussian QI, our protocol, and the CI by using the QCB which assumes the optimal receiver. The quantum advantage is analyzed with the amount of the quantum discord. In Sec.~\ref{SecROC} a realistic receiver is considered in those protocols, and finally it is summarized with discussion in Sec.~\ref{SecCon}.

\section{Quantum illumination with asymmetrically squeezed two-mode state}\label{SecASTMQI}

\begin{figure}[t!]
	\centering	
	\includegraphics[width=0.4\textwidth]{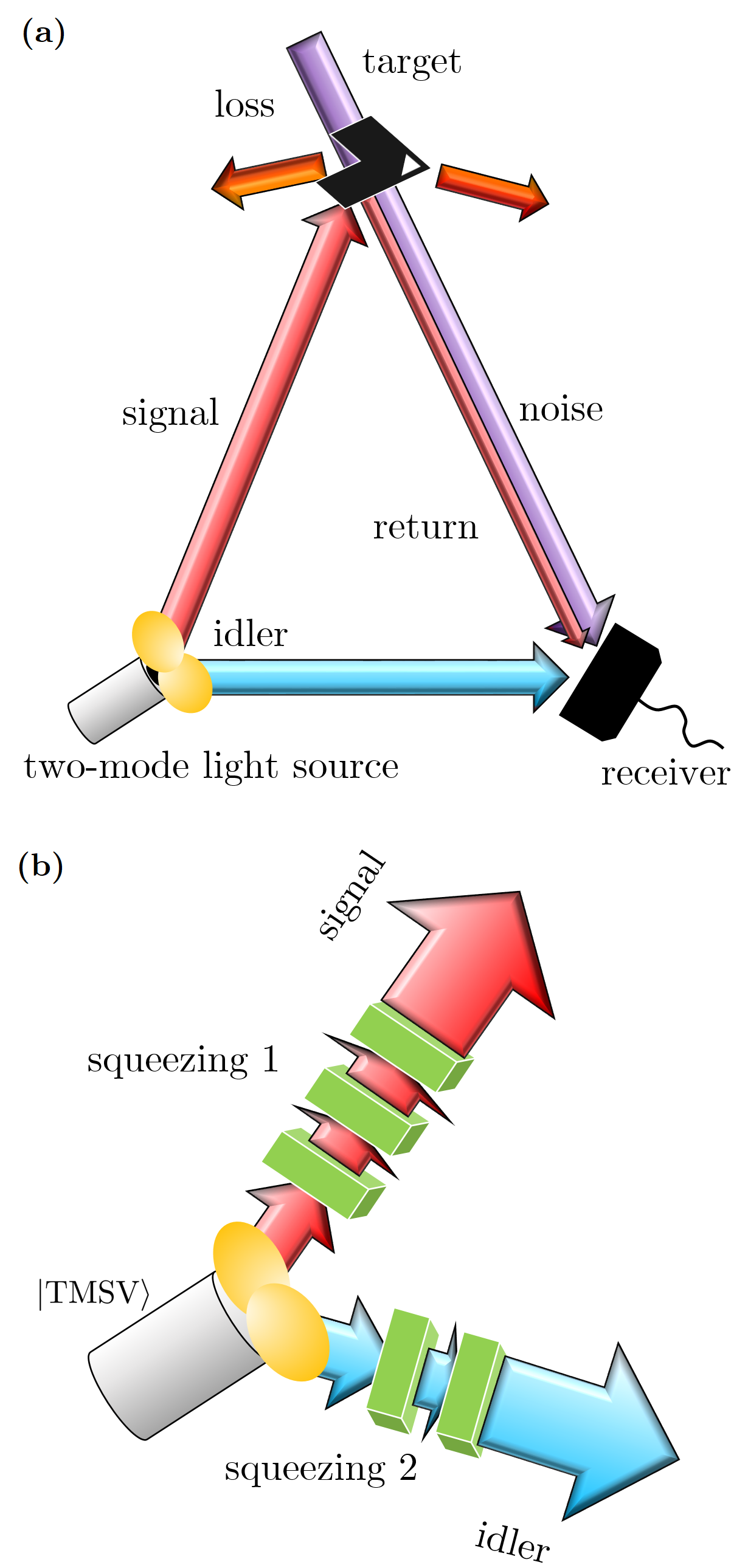}
	\caption{(a) A schematic diagram of QI, where an entangled state is exploited to discriminate the two situations, target presence/absence. The signal mode is sent to the target, and the idler mode is retained. The return mode which includes the reflected signal and a thermal noise is jointly measured with the idler mode for decision of target existence. (b) Generation of an ASTM state, where each mode of the initial TMSV state is independently squeezed.}\label{FigQI}
\end{figure}

Fig.~\ref{FigQI}(a) shows a schematic diagram of QI which is a remote target detection protocol that exploits an entangled state. The signal mode of the entangled state is sent to a target, and the idler mode is ideally retained. If the target presents, the return mode which includes the reflected signal and a background thermal noise is jointly measured with the idler mode. If there is no target, the receiver only gets the background noise with the idler mode. On the other hand, its classical counterpart exploits a single-mode light to discriminate the two situations with no idler mode. Since the optimal CI strategy exploits a coherent state \cite{SL09}, QI should have a lower error rate of the discrimination than the CI with the same energy to claim the quantum advantage.

\subsection{ASTM state}
A conventional Gaussian QI protocol exploits a TMSV state which is written as
\begin{align}
	\ket{\text{TMSV}}=\sum_{n=0}^{\infty}\sqrt{\frac{(N_{0})^{n}}{(N_{0}+1)^{n+1}}}\ket{n}_{1}\ket{n}_{2},
\end{align}
where $N_{0}$ is the mean photon number of each mode, and the subscripts $1$ and $2$ denote two different modes. Single-mode and two-mode squeezing operations are defined as follows:
\begin{align}
	\begin{split}
		\hat{S}_{i}(z)&=\exp\left[\frac{1}{2}\left(z^{\ast}\hat{a}^{2}-z\hat{a}^{\dagger 2}\right)\right],\\
		\hat{S}_{12}(z)&=\exp\left(z^{\ast}\hat{a}_{1}\hat{a}_{2}-z\hat{a}_{1}^{\dagger}\hat{a}_{2}^{\dagger}\right),
	\end{split}
\end{align}
where $z=r e^{i\phi}$ with squeezing parameter $r$ and squeezing angle $\phi$. $\hat{a}^{\dagger}$ and $\hat{a}$ are photon creation and annihilation operators. The mean photon number of a squeezed vacuum state is given by $N=\sinh^{2} r$ \cite{Weedbrook2012}.

Intuitively, a strong signal would give better sensitivity in a remote target detection protocol. However, there is a limit to increase the squeezing parameter under current technologies. The squeezing parameters in recent demonstrations are roughly $1.73$($N\approx 7.41$) in optical frequency \cite{Vahlbruch2016} and $0.944$($N\approx 1.19$) in a microwave domain \cite{Dassonneville2021}. Thus, we apply local squeezing operations on a TMSV state in order to increase the signal energy. In Fig.\ref{FigQI}(b), local squeezing operations are performed on the signal and idler modes of the TMSV state. We call it an asymmetrically squeezed two-mode(ASTM) state. 

For simplicity, we assume $\phi_{1}=\phi_{2}=0$ for all additional squeezing operations. This assumption does not change the performance of QI under the QCB, where the relative squeezing angle between the signal and idler modes is optimized at the receiver. Under a realistic receiver, the optimal condition becomes $\phi_{1}=\phi_{2}$ which will be shown in Sec.~\ref{SecROC}.

For Gaussian states, it is convenient to use quadrature representations. An ASTM state is given by
\begin{align}
	\mathbf{V}_{\text{ASTM}}=\mathcal{M}_{S}\mathbf{V}_{\text{TMSV}}\mathcal{M}_{S}^{T},
\end{align}
where the covariance matrix of a TMSV state is
\begin{align}
	\mathbf{V}_{\text{TMSV}}=\begin{pmatrix}
		A & 0 & C & 0\\
		0 & A & 0 & -C\\
		C & 0 & A & 0\\
		0 & -C & 0 & A
	\end{pmatrix},
\end{align}
with $A=2N_{0}+1$ and $C=2\sqrt{N_{0}(N_{0}+1)}$. The symplectic matrix $\mathcal{M}_{S}$ corresponding to $\hat{S}_{1}(r_{1})\hat{S}_{2}(r_{2})$ is written as
\begin{align}
	\mathcal{M}_{S}=\begin{pmatrix}
		\gamma_{1-} & 0 & 0 & 0\\
		0 & \gamma_{1+} & 0 & 0\\
		0 & 0 & \gamma_{2-} & 0\\
		0 & 0 & 0 & \gamma_{2+}
	\end{pmatrix},
\end{align}
with $\gamma_{j\pm}=\sqrt{N_{j}+1}\pm\sqrt{N_{j}}$ and $N_{j}=\sinh^{2}r_{j}$ \cite{Weedbrook2012}. The mean photon numbers of the signal and idler modes in the ASTM state are $N_{S}=N_{0}+2N_{0}N_{1}+N_{1}$ and $N_{I}=N_{0}+2N_{0}N_{2}+N_{2}$, respectively. Since $N_{j}$ with $j\in\{1,2\}$ is a non-negative value, the mean photon number of the signal and idler modes increases as high as $(1+2N_{0})N_{j}$.

In QI, a target can be modeled by a beam splitter with a reflectivity $\kappa$, where the signal mode of an ASTM state is injected into one of the input ports, and a thermal noise into the other input port. One of the output ports, which includes the reflected signal and the transmitted thermal noise, is directed to a receiver whereas the other output port is discarded to simulate a loss channel. Since the mean photon number of the thermal noise is defined as $N_{B}/(1-\kappa)$, the mean photon number of the thermal noise at the receiver is always $N_{B}$ regardless of the target reflectivity. If the target exists, the covariance matrix of the output mode is written as follows:
\begin{align}\label{state}
	\begin{split}
	\mathbf{V}_{A}=&\begin{pmatrix}
		\sqrt{\kappa}\mathbf{I}_{2} & \mathbf{0}_{2}\\
		\mathbf{0}_{2} & \mathbf{I}_{2}
	\end{pmatrix}\mathbf{V}_{\text{ASTM}}\begin{pmatrix}
		\sqrt{\kappa}\mathbf{I}_{2} & \mathbf{0}_{2}\\
		\mathbf{0}_{2} & \mathbf{I}_{2}
	\end{pmatrix}\\
	&+(1-\kappa)\begin{pmatrix}
		2\frac{N_{B}}{1-\kappa}+1 & 0 & 0 & 0\\
		0 & 2\frac{N_{B}}{1-\kappa}+1 & 0 & 0\\
		0 & 0 & 0 & 0\\
		0 & 0 & 0 & 0
	\end{pmatrix},
	\end{split}
\end{align}
and if there is no target $(\kappa=0)$, the covariance matrix becomes
\begin{align}
	\mathbf{V}_{B}=\begin{pmatrix}
		2N_{B}+1 & 0 & 0 & 0\\
		0 & 2N_{B}+1 & 0 & 0\\
		0 & 0 & \gamma_{2-}^{2}A & 0\\
		0 & 0 & 0 & \gamma_{2+}^{2}A
	\end{pmatrix},
\end{align}
where $\mathbf{I}_{2}$ is the $2\times 2$ identity matrix and $\mathbf{0}_{2}$ denotes the $2\times 2$ null matrix. Note that the quantum states we consider, a squeezed vacuum state and a thermal state, are zero-mean Gaussian states, so that we do not need to consider the first-order moments.

\subsection{Performance}

In a target detection problem, there are error probabilities in the binary decision: target presence/absence. The total error is written in the following equation:
\begin{align}
P_{E}=P(0)P(1|0)+P(1)P(0|1),
\end{align}
where $P_{E}$ is a decision error probability, $P(x)$ denotes the prior probability associated with the hypothesis $x$, $P(y|x)$ is the probability that $y$ is chosen in the hypothesis of $x$. $1$ and $0$ denote the target presence and absence, respectively. The probability of type-I error, false alarm, is $P(1|0)$, and that of type-II error, missed detection, is $P(0|1)$. With these error probabilities, we can investigate the performance of a target detection.

\section{Optimal QI receiver}\label{SecPA}
\subsection{Quantum Chernoff bound}

If we have no prior information about target presence, it is natural to assume a random event, $P(0)=P(1)=1/2$. In this case, it is necessary to minimize the total error probability, and the QCB provides an upper bound to the minimum error probability of the statistical discrimination problem of two quantum states \cite{Calsamiglia2008}. The definition of the QCB is as follows \cite{Audenaert2007}:
\begin{align}
	P_{E}\leq P_{\text{QC}}^{(M)}=\frac{1}{2}\left[\inf_{0\leq s\leq 1}Q_{s}\right]^{M},
\end{align}
where $P_{\text{QC}}^{(M)}$ denotes the QCB, $M$ is the number of ensembles for the decision, and
\begin{align}\label{Qs}
	Q_{s}\coloneqq \text{Tr}(\rho_{A}^{s}\rho_{B}^{1-s}).
\end{align}
The density matrices $\rho_{A}$ and $\rho_{B}$ are the quantum states.

The QCB of Gaussian states can be calculated by first-order moments and symplectic spectrum of the two covariance matrices. For every $4\times 4$ covariance matrix, there exists a symplectic matrix $\mathbf{S}$ that satisfies
\begin{align}
	\mathbf{V}=\mathbf{S}\left[\bigoplus_{k=1}^{2}\nu_{k}\mathbf{I}_{2}\right]\mathbf{S}^{T},
\end{align}
where $\nu_{k}$ is a symplectic eigenvalue and $\mathbf{I}_{2}$ denotes a $2\times 2$ identity matrix. To describe the QCB, we define the following equations:
\begin{align}
	\begin{split}
		G_{p}(x)&\coloneqq\frac{2^{p}}{(x+1)^{p}-(x-1)^{p}},\\
		\Lambda_{p}(x)&\coloneqq \frac{(x+1)^{p}+(x-1)^{p}}{(x+1)^{p}-(x-1)^{p}},\\
		\mathbf{V}(p)&\coloneqq \mathbf{S}\left[\bigoplus_{k=1}^{2}\Lambda_{p}(\nu_{k})\mathbf{I}_{2}\right]\mathbf{S}^{T}.
	\end{split}
\end{align}
Then, Eq.~(\ref{Qs}) can be calculated from the following equations \cite{Pirandola2008}:
\begin{align}
	Q_{s}=\widetilde{Q}_{s}\exp\left\{-\frac{1}{2}\vec{d}^{T}\left[\mathbf{V}_{A}(s)+\mathbf{V}_{B}(1-s)\right]^{-1}\vec{d}\right\},
\end{align}
where
\begin{align}\label{Qst}
	\widetilde{Q}_{s}\coloneqq \frac{4 \prod_{k=1}^{2}G_{s}(\alpha_{k})G_{1-s}(\beta_{k})}{\sqrt{\det\left[\mathbf{V}_{A}(s)+\mathbf{V}_{B}(1-s)\right]}}.
\end{align}
with $\vec{d}=\vec{x}_{A}-\vec{x}_{B}$ and
\begin{align}\nonumber
\vec{x}^{T}\coloneqq (\braket{\hat{x}_{R}},\braket{\hat{p}_{R}},\braket{\hat{x}_{I}},\braket{\hat{p}_{I}}).
\end{align}
The subscripts $R$ and I denote the return and idler modes, respectively. Using $\alpha_{k}$($\beta_{k}$), the symplectic eigenvalues of $\mathbf{V}_{A}$($\mathbf{V}_{B}$), we can calculate the denominator in Eq.~(\ref{Qst}) with
\begin{align}
	\mathbf{V}_{A}(s)&=\mathbf{S}_{A}\left[\bigoplus_{k=1}^{2}\Lambda_{s}(\alpha_{k})\mathbf{I}_{2}\right]\mathbf{S}_{A}^{T},\\
	\mathbf{V}_{B}(1-s)&=\mathbf{S}_{B}\left[\bigoplus_{k=1}^{2}\Lambda_{1-s}(\beta_{k})\mathbf{I}_{2}\right]\mathbf{S}_{B}^{T}.
\end{align}
Under zero first-order moment, we obtain $Q_{s}=\widetilde{Q}_{s}$.

Here, we analyze the performance of QI by using the signal-to-noise ratio(SNR) that is related to the total error probability in the following equation \cite{Karsa2020-1,Barzanjeh2020,Jo2021}:
\begin{align}
	P_{QC}^{(M)}=\frac{1}{2}\text{erfc}\left[\sqrt{\text{SNR}^{(M)}_{\text{QC}}}\right],
\end{align}
where $\text{erfc}(x)$ is the complementary error function. From the functional form of $\text{erfc}(x)$, a lower QCB implies a higher SNR. By omitting the subscript QC and the superscript $(M)$, the SNR indicates $\text{SNR}_{\text{QC}}^{(M)}$ from now on.

\subsection{Performance analysis}

We analyze the SNR of the Gaussian QI under additional squeezing operations on the each mode of the initial TMSV state. Since we are interested in microwave regime to detect a long distance target, initially we set the mean photon number of the initial TMSV state as $N_{0}=1$ \cite{Sandbo,Barzanjeh2020} and the energy of the thermal noise as $N_{B}=3.8\times 10^{3}$ which corresponds to a thermal radiation of a few GHz frequency at the room temperature. The target reflection is fixed as $\kappa=0.01$, and the number of ensembles is $M=10^{7}$.

\begin{figure}[ht!]
	\centering	
	\includegraphics[width=0.45\textwidth]{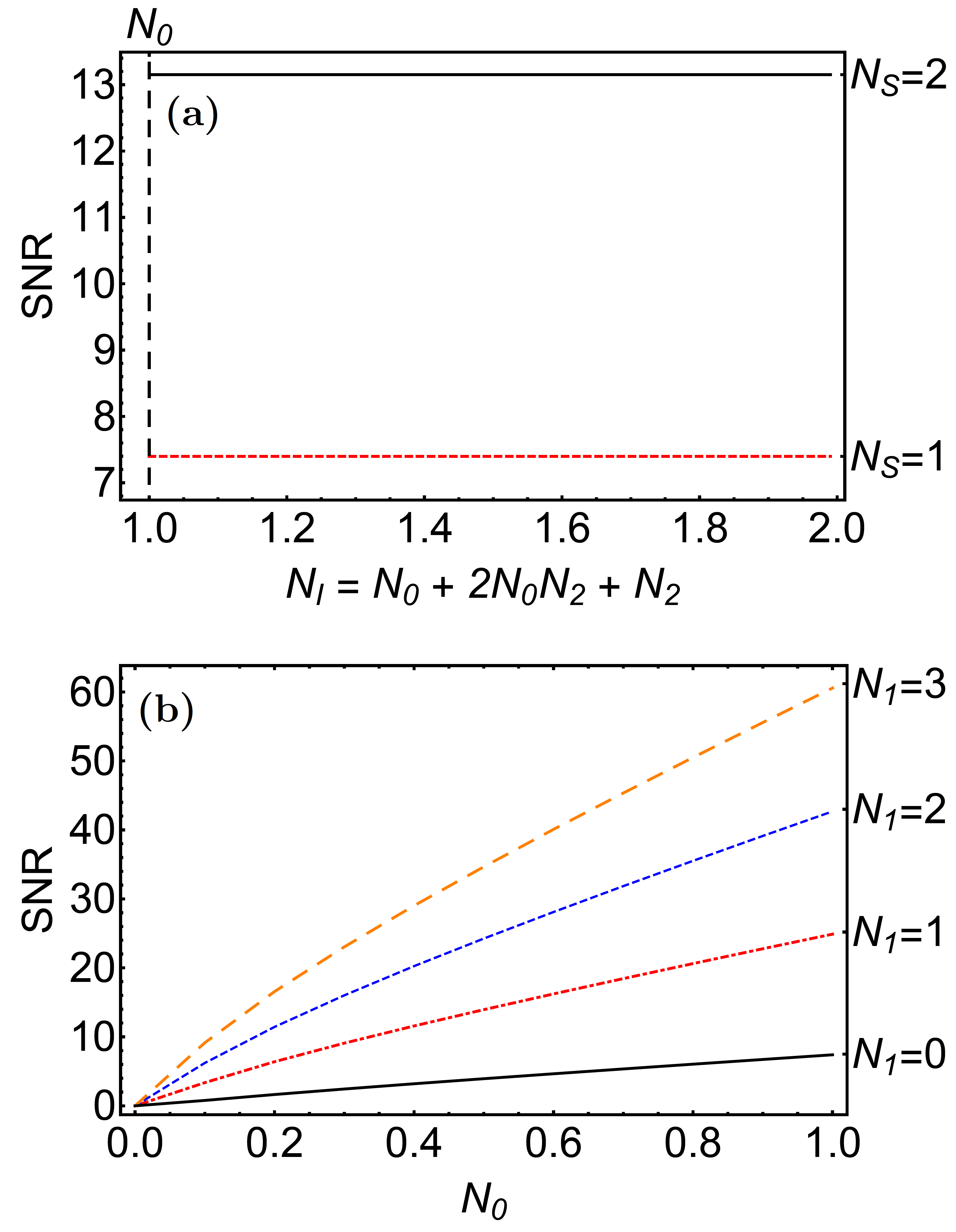}
	\caption{(a) Given a fixed signal mode $N_{S}$, SNR as a function of the idler mode $N_{I}=N_{0}+2N_{0}N_{2}+N_{2}$, where $N_{2}=\sinh^{2}r_{2}$. The red dashed line denotes the case $N_{S}=1$, and the black line does $N_{S}=2$. (b) Given a fixed idler mode $N_{I}=N_{0}$, SNR as a function of the signal mode of the initial TMSV state $N_{0}$ with the additional squeezing parameter $N_{1}$. The other parameter values are $N_{0}=1$, $N_{B}=3.8\times 10^{3}$, $\kappa=0.01$, and $M=10^{7}$.}\label{FigIdler}
\end{figure}

Fig.~\ref{FigIdler}(a) shows the SNRs in the case that the idler mode energy $N_{I}=N_{0}+2N_{0}N_{2}+N_{2}$ is amplified from $N_{0}$. The red dashed line denotes the SNR when there is no additional squeezing on the signal mode($N_{S}=N_{0}=1$), and the black line does the SNR with the additionally squeezed signal mode $N_{S}=2$. Both plots are not changed by varying $N_{I}$ implying the SNR is not affected by the additional squeezing operation on the idler mode. Since we analyze the performance of QI by using the QCB that assumes the optimal measurement, any additional single-mode operation on the idler mode before the measurement is included in the optimal receiver. Therefore, an additional squeezing on the idler mode does not affect the SNR based on the QCB.

In Fig.~\ref{FigIdler}(b), we show the SNRs of QI using an ASTM state with an additional squeezing parameter $N_{1}$, against the energy of the initial TMSV state. Recall that the signal energy of the ASTM state is given by $N_{S}=N_{0}+2N_{0}N_{1}+N_{1}$. Starting with the TMSV state of $N_{S}=N_{0}=1$, which was experimentally reported in a microwave frequency \cite{Sandbo}, we obtain $\text{SNR}\approx 7$ at $N_{1}=0$. The SNR increases with the additional squeezing $N_{1}$; $\text{SNR}\approx 25$, $43$, $61$ at $N_{1}=1$, $2$, $3$, respectively. Thus, we can claim that the additional single-mode squeezing on the signal mode provides a significant improvement in the Gaussian QI. The enhancement can be explained by the signal mode energy as well as the cross-correlation terms, $\braket{\hat{a}_{R}\hat{a}_{I}}$ and $\braket{\hat{a}_{R}^{\dagger}\hat{a}_{I}^{\dagger}}$, which are exploited to detect a target \cite{Guha2009,Jo2021}. The single-mode squeezing operation on the signal mode amplifies the signal energy from $N_{0}$ to $N_{0}+2N_{0}N_{1}+N_{1}$ also the cross-correlation from $\sqrt{N_{0}(N_{0}+1)}$ to $\sqrt{N_{0}(N_{0}+1)(N_{1}+1)}$. These enlarged signal energy and cross-correlation lead to the performance improvement in the Gaussian QI.

\begin{figure}[t!]
	\centering	
	\includegraphics[width=0.45\textwidth]{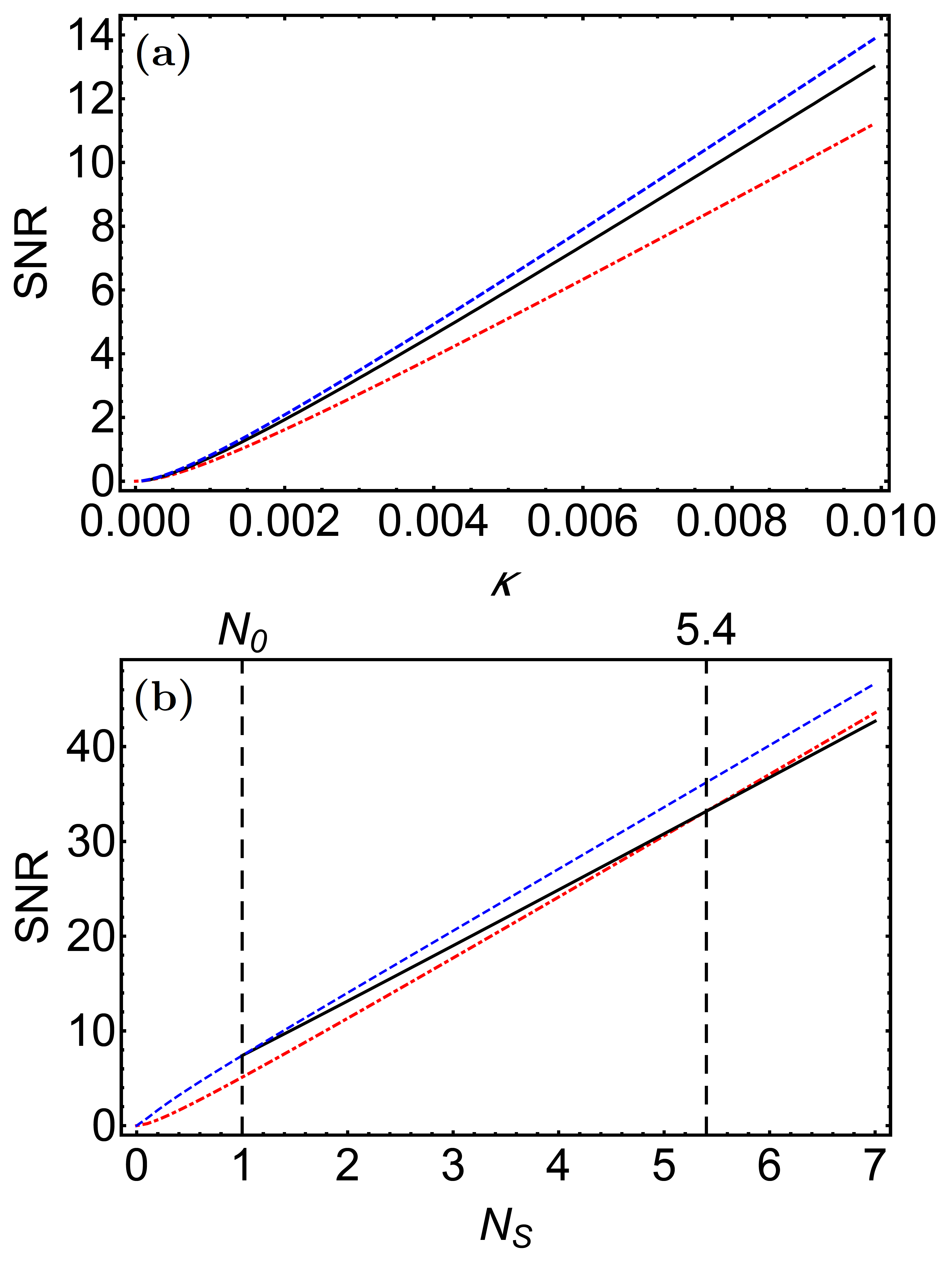}
	\caption{SNR using a bright TMSV state($N_{S}=N_{0}$, blue dashed lines), a coherent state($N_{S}=|\alpha|^{2}$, red dot-dashed lines), and an ASTM state ($N_{S}=N_{0}+2N_{0}N_{1}+N_{1}$, black lines). (a) Given a fixed $N_{S}=2$, SNR as a function of the target reflectivity $\kappa$. The ASTM state is generated from the initial TMSV state with $N_{0}=1$. (b) Given a fixed target reflectivity $\kappa=0.01$, SNR as a function of the total input energy in the signal mode $N_{S}$. The other parameter values are the same as those used in Fig.~\ref{FigIdler}.}\label{FigQICI}
\end{figure}

Given a fixed transmitted energy, we compare an ASTM state with a bright TMSV one in the performance of QI, while a coherent state is considered for a classical benchmark. The signal energies of a TMSV state, a coherent state, and an ASTM state are given by $N_{S}=N_{0}$, $|\alpha|^{2}$, and $N_{0}+2N_{0}N_{1}+N_{1}$, respectively. The ASTM state is generated from the initial TMSV state with $N_{0}=1$. Since a squeezing on the idler mode does not change the performance of QI, only the signal mode is additionally squeezed. Fig.~\ref{FigQICI}(a) shows the SNRs as a function of the target reflectivity, where the signal energy is fixed as $N_{S}=2$ for all the states. It shows the SNR with the ASTM state is close to that using the bright TMSV state under the same signal energy, for example, the ratio at $\kappa=0.05$ is $93.4$\%. Moreover, our protocol always provides the quantum advantage in the plotted region. Therefore, if the bright TMSV state is not available, we can replace it by the ASTM state with the same signal energy.

In Fig.~\ref{FigQICI}(b), the SNR with an ASTM state shows better performance than the CI at $N_{S}\leq 5.4$, while it is competitive with the performance of the conventional Gaussian QI with the same signal energy. For example, the SNR of our protocol is $93.8$\% ($N_{S}=2$) and $91.9$\% ($N_{S}=4$) of the SNR using a TMSV state with the same energy. The result suggests that our QI protocol is an implementable alternative rather than generating a microwave TMSV state with the same energy. 

\begin{figure}[t!]
	\centering	
	\includegraphics[width=0.45\textwidth]{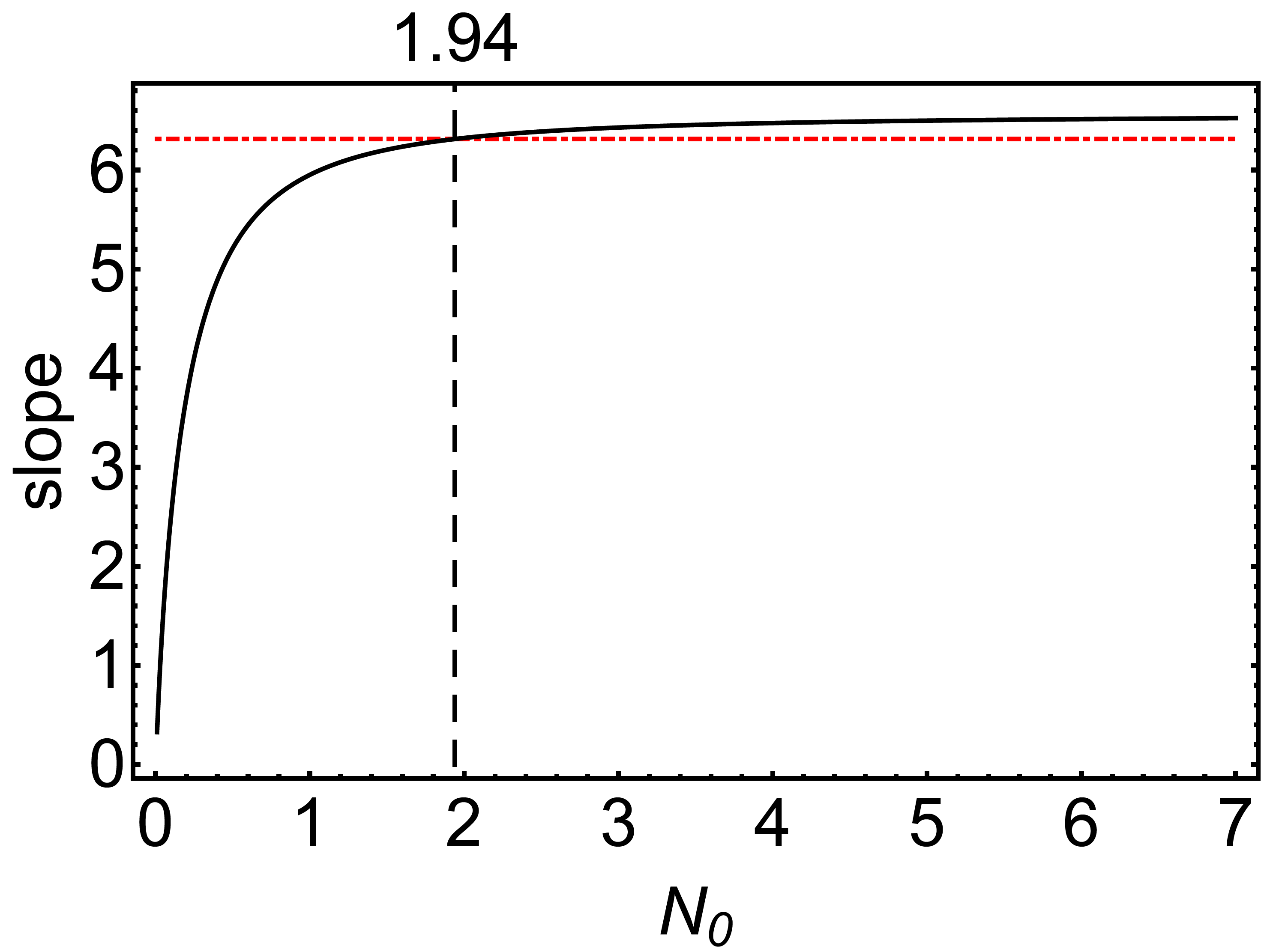}
	\caption{Approximated slope of the SNR using an ASTM state as a function of the initial input energy in the signal mode $N_{0}$ (black line). The red dot-dashed line denotes the slope of the CI. The parameter values are the same as those of Fig.~\ref{FigQICI}(b).}\label{FigLimit}
\end{figure}

However, our protocol does not always provide quantum advantage over the CI, as shown in Fig.~\ref{FigQICI}(b). Since the slope of the SNR with the ASTM state is smaller than that of the CI, the quantum advantage of our protocol vanishes with a large additional squeezing. If the slope is equal to or larger than that of the CI, QI with the ASTM state always has better performance than the CI. In Fig.~\ref{FigLimit}, we show that the approximated slope of the SNR with the ASTM state mainly depends on the initial signal energy, leading to that the initial TMSV state should have energy $N_{0}\geq 1.94$ in order to beat the CI. This limit has not reached yet under microwave technologies.

\subsection{Quantum advantage originated from quantum discord?}\label{SecQAQD}

Different from other quantum information protocols, QI takes advantages over CI even when entanglement is broken before the measurement process. Since quantum discord exists without entanglement \cite{Ollivier2001,Henderson2001,Modi2012}, there was a claim that quantum discord can be the resource of quantum advantage of QI in discrete \cite{Weedbrook} and Gaussian \cite{Bradshaw2017} systems. However, we show that quantum discord cannot fully characterize quantum advantages of Gaussian QI. It was recently mentioned with discrete signals \cite{Yung}.

We start from the covariance matrix of a two-mode Gaussian state:
\begin{align}
	\mathbf{V}=\begin{pmatrix}
		V_{11} & V_{12}\\
		V_{21} & V_{22}
	\end{pmatrix},
\end{align}
where $V_{ij}$ is a $2\times 2$ block matrix, and we define $\alpha\coloneqq \text{Det}[V_{11}]$, $\beta\coloneqq \text{Det}[V_{22}]$, $\gamma\coloneqq \text{Det}[V_{12}]=\text{Det}[V_{21}]$, and $\delta\coloneqq\text{Det}[\mathbf{V}]$. Then the quantum discord for a Gaussian two-mode state can be calculated from the following equation \cite{Giorda2010,Adesso2010}:
\begin{align}\label{QD}
	\text{QD}=f(\sqrt{\beta})-f(\sqrt{\nu_{-}})-f(\sqrt{\nu_{+}})+f(\sqrt{\epsilon}),
\end{align}
where
\begin{align}
	f(x)&\coloneqq\frac{x+1}{2}\ln\frac{x+1}{2}-\frac{x-1}{2}\ln\frac{x-1}{2},\\
	\nu_{\pm}&\coloneqq\alpha+\beta+2\gamma \pm \sqrt{\frac{(\alpha+\beta+2\gamma)^2-4\delta}{2}},
\end{align}
and
\begin{widetext}
\begin{align}\label{eps}
\epsilon \coloneqq \left\{\begin{matrix}
\dfrac{2\gamma^{2}+(\beta-1)(\delta-\alpha)+2|\gamma|\sqrt{\gamma^{2}+(\beta-1)(\delta-\alpha)}}{(\beta-1)^2} & \text{if }(\delta-\alpha\beta)^{2}\leq (\beta+1)\gamma^{2}(\alpha+\delta),\\ & \\
\dfrac{\alpha \beta-\gamma^{2}+\delta-\sqrt{\gamma^{4}+(\delta-\alpha\beta)^{2}-2\gamma^{2}(\delta+\alpha\beta)}}{2\beta} & \text{otherwise.}
\end{matrix}\right.
\end{align}
\end{widetext}

\begin{figure}[t!]
	\centering	
	\includegraphics[width=0.45\textwidth]{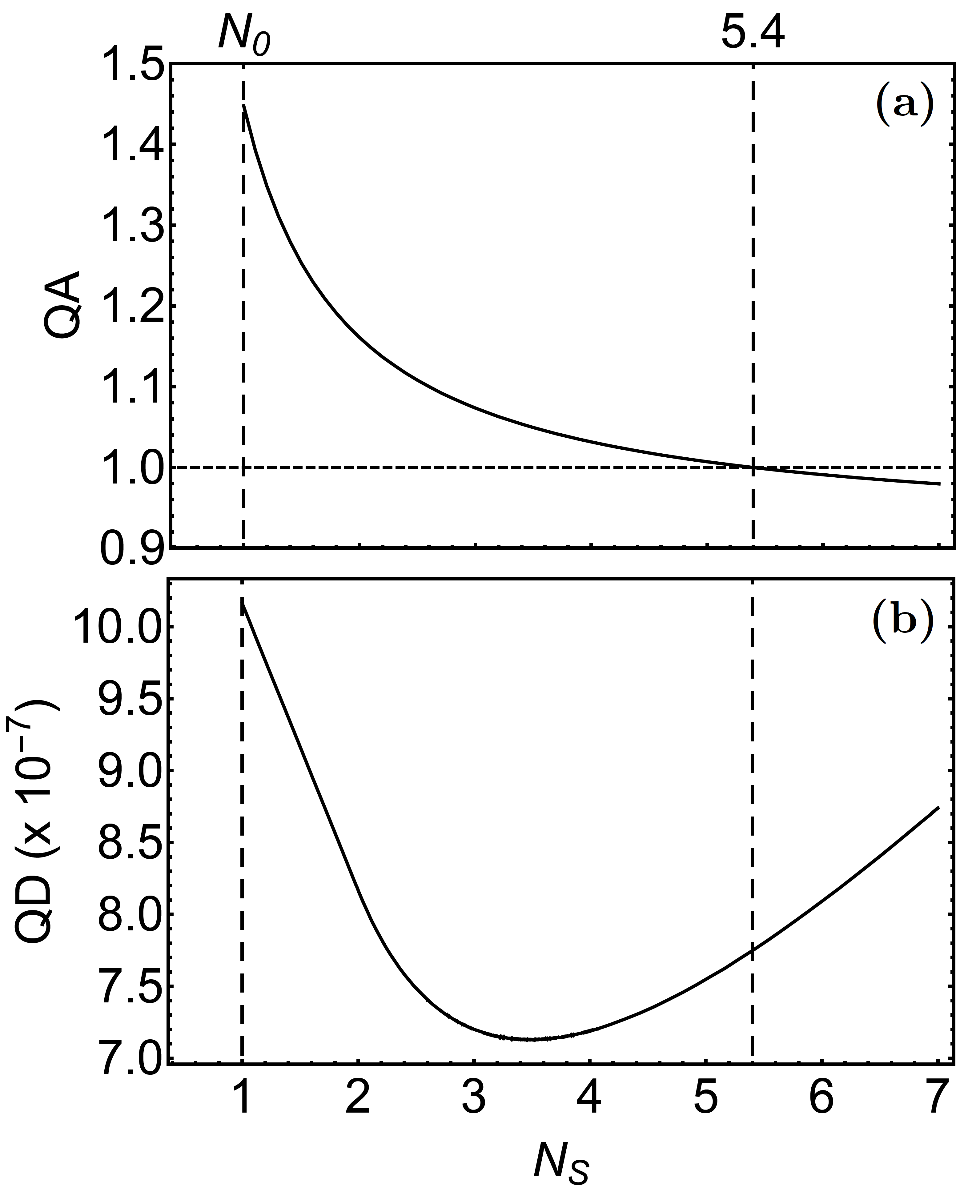}
	\caption{(a) Quantum advantage(QA) of QI using an ASTM state over the CI as a function of the total input energy in the signal mode $N_{S}=N_{0}+2N_{0}N_{1}+N_{1}$. (b) Quantum discord(QD) between the return and idler modes of the ASTM state as a function of $N_{S}$. The parameter values are the same as those of Fig.~\ref{FigQICI}(b).}\label{FigQAQD}
\end{figure}

In Eqs.~(\ref{QD}--\ref{eps}), the quantum discord is calculated from the determinants of the block matrices of the covariance matrix. Since the additional squeezing operations for generating an ASTM state are local operations, they do not change the determinants. Thus, the quantum discord of the ASTM state is the same as that of the initial TMSV state. However, we find that the remaining quantum discord after the channel, i.e., the quantum discord between the return and idler modes, is changed with the additional squeezing. We compare the remaining quantum discord with the performance of QI. Fig.~\ref{FigQAQD}(a) shows the quantum advantage of our QI compared to the CI. The quantum advantage is calculated from $\text{SNR}_{\text{ASTM}}/\text{SNR}_{\text{CI}}$ under the same energy, where the both SNRs are already shown in Fig.~\ref{FigQICI}(a). Compared to the observation that the quantum advantage is a monotonic function with increasing $N_{S}$, the quantum discord between the return and idler modes is not a monotonic function of $N_{S}$, as shown in Fig.~\ref{FigQAQD}(b). Moreover, the quantum advantage monotonically decreases with increasing $N_{S}$, whereas the quantum discord increases at roughly $N_{S}\geq 3.4$, leading to the opposite behavior at large $N_{S}$. Therefore, we can claim that the quantum discord cannot be the resource of the quantum advantage in the Gaussian QI.

\section{Realistic QI receiver}\label{SecROC}
\subsection{Receiver operating characteristic}

In this section, we investigate the performance of QI with a double heterodyne detection(dHTD) receiver, of which the observable form is written as follows:
\begin{align}\label{dHet}
\hat{O}_{\text{dHTD}}=\hat{a}_{R}^{\dagger}\hat{a}_{I}^{\dagger}+\hat{a}_{R}\hat{a}_{I}.
\end{align}
The dHTD receiver is the nearly optimal observable in Gaussian QI \cite{Lee2021}.

With the dHTD receiver, we analyze receiver operating characteristic(ROC) that illustrates the performance of a target detection by plotting missed detection probability $P(0|1)$ against false alarm probability $P(1|0)$. Given various decision thresholds, we obtain two more probabilities, such as detection probability $P(1|1)$ and null probability $P(0|0)$. Under the conditions of $P(0|1)+P(1|1)=1$ and $P(1|0)+P(0|0)=1$, we investigate $P(0|1)$ and $P(1|0)$ in a single plot. The best occasion is that both probabilities $P(0|1)$ and $P(1|0)$ converge to zero. There are three regions in the plot: a good region at $1-P(0|1) > P(1|0)$, random at $1-P(0|1)=P(1|0)$, and otherwise a bad region.

Given many ensembles, we can construct Gaussian distributions from measurement outcomes of the receiver. The corresponding mean $M R_{x}$ and variance $M \Delta R_{x}$ are calculated by
\begin{align}\label{meanvar}
\begin{split}
R_{x}&=\text{Tr}\left[\hat{O}_{\text{dHTD}}\hat{\rho}_{x}\right] ,\\
\Delta R_{x}&=\text{Tr}\left[ (\hat{O}_{\text{dHTD}})^{2}\hat{\rho}_{x} \right] -\left(\text{Tr}\left[\hat{O}_{\text{dHTD}}\hat{\rho}_{x}\right]\right)^{2},
\end{split}
\end{align}
where $\hat{\rho}_{x}$ is a quantum state corresponding to the covariance matrix $\mathbf{V}_{x}$, $x\in\{A, B\}$, and $M$ is the number of ensembles. The analytic forms of the mean and variance are given in Appendix \ref{SecApp1}. Under the condition of $R_{A}\geq R_{B}$ and $M\gg 1$, we obtain the two error probabilities \cite{Jo2021}:
\begin{align}
\begin{split}
P_{\text{FA}}&\coloneqq P(1|0)=\frac{1}{2}\text{erfc}\left[\frac{R_{\text{Th}}-M R_{B}}{\sqrt{2 M \Delta R_{B}}}\right],\\
P_{\text{MD}}&\coloneqq P(0|1)=\frac{1}{2}\text{erfc}\left[\frac{M R_{A}-R_{\text{Th}}}{\sqrt{2 M \Delta R_{A}}}\right],
\end{split}
\end{align}
where $R_{\text{Th}}$ is the decision threshold, and erfc$(z)\coloneqq 1-2 \pi^{-1/2} \int_{0}^{z} \exp(-t^{2})dt$ is the complementary error function. Then, we can obtain a ROC curve by scanning $R_{\text{Th}}$ from $-\infty$ to $\infty$.

\subsection{Performance analysis}

\begin{figure}[t!]
	\centering	
	\includegraphics[width=0.42\textwidth]{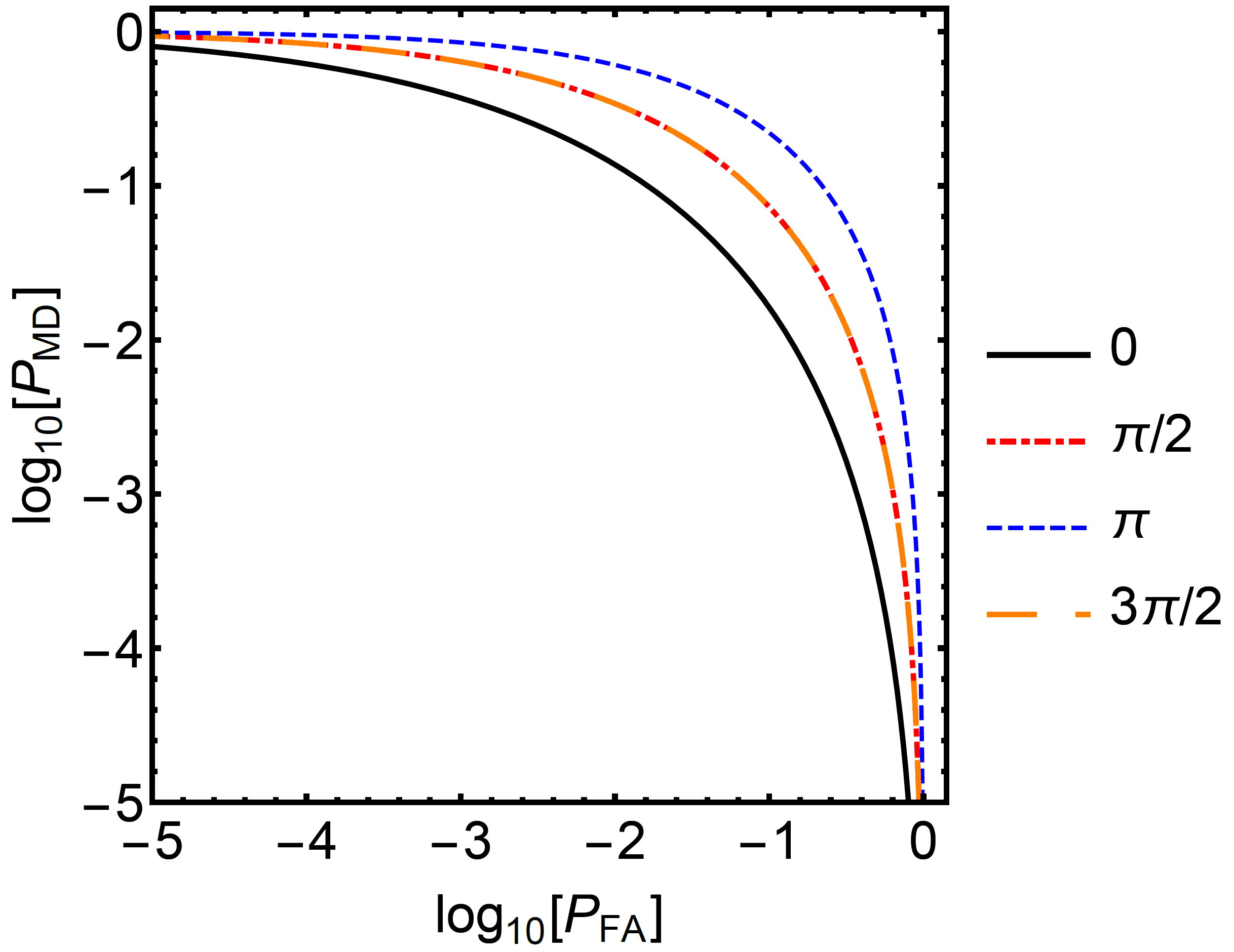}
	\caption{ROC curves of QI using an ASTM state under various relative angles of the two local squeezing operations, $\phi_{1}-\phi_{2}$. The performance is the best at $\phi_{1}=\phi_{2}$, and the two 		ROC curves at $\phi_{1}-\phi_{2}=\pi/2$ and $3\pi/2$ are overlapped.The other parameters are given as $N_{0}=1$, $N_{S}=N_{I}=2$, $N_{B}=3.8 \times 10^{3}$, $\kappa=0.01$, and $M=10^{6}$.}\label{FigAng}
\end{figure}

Fig.~\ref{FigAng} shows the effect of local squeezing angle in QI with an ASTM state. Since lowering the error probabilities represent better performance, converging to the left-bottom corner exhibits better performance in the ROC curve. In the figure, QI with the ASTM state has the best performance at the two squeezing angles $\phi_{1}=\phi_{2}$, regardless of the local squeezing parameters $N_{1}$ and $N_{2}$. It supports that our assumption of $\phi_{1}=\phi_{2}=0$ is reasonable even in the dHTD receiver.

\begin{figure}[t!]
	\centering	
	\includegraphics[width=0.45\textwidth]{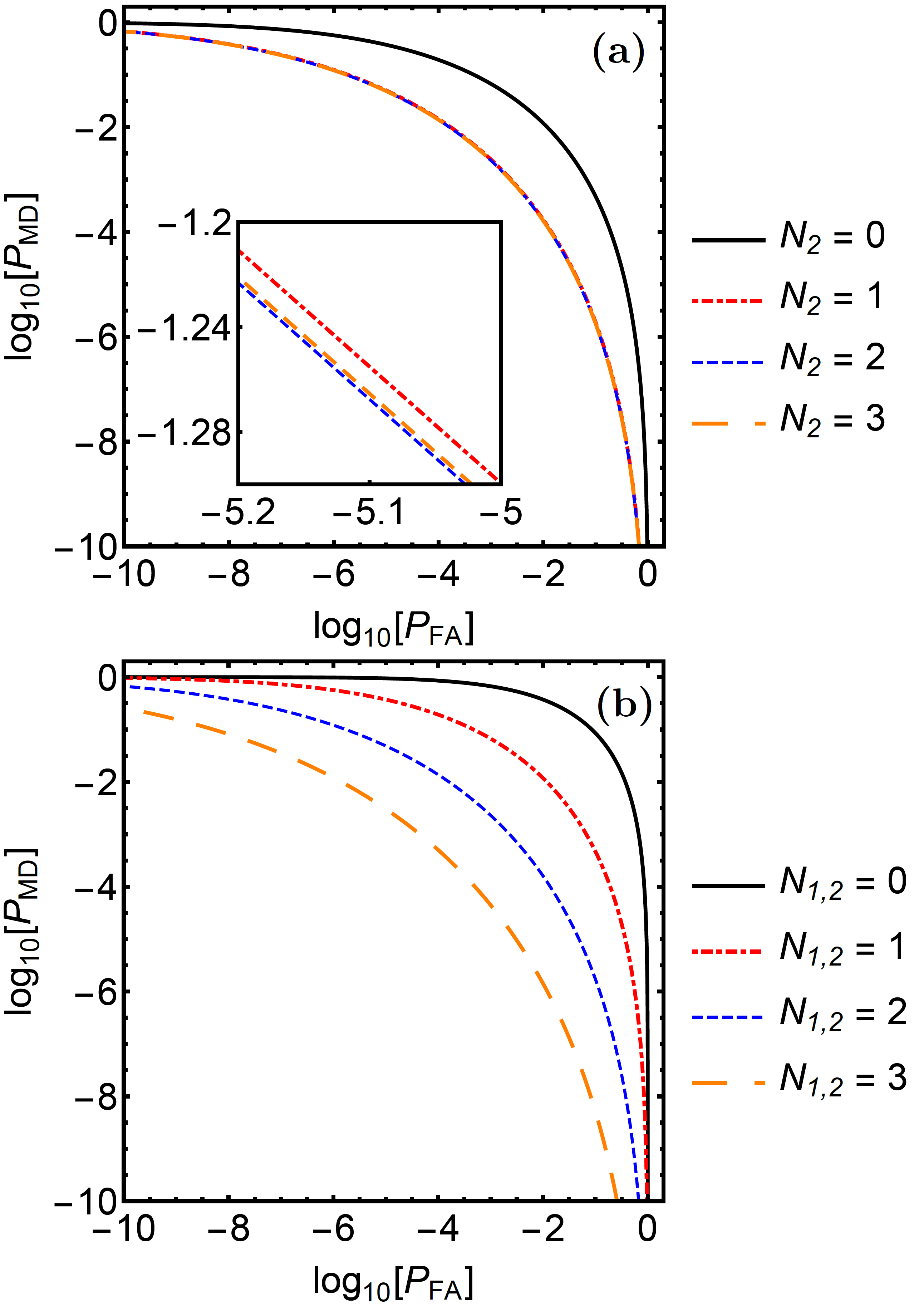}
	\caption{ROC curves of QI with an ASTM state under various local squeezing parameters. (a) ROC curves with various squeezing parameters on the idler mode $N_{2}$ at $N_{1}=2$. The performance is the best at $N_{1}=N_{2}$ among the plots. (b) ROC curves with equal local squeezing parameter $N_{1}=N_{2}\coloneqq N_{1,2}$. The other parameter values in the plots are the same as those used in Fig.~\ref{FigAng}.}\label{FigROC}
\end{figure}

At $\phi_{1}=\phi_{2}=0$, we analyze the effect of local squeezing parameters in the QI. For a fixed squeezing parameter of $N_{1}=2$ in the signal mode, we plot the ROC curves under various squeezing parameters $N_{2}$ in the idler mode, as shown in Fig.~\ref{FigROC} (a). It shows that the best performance is obtained when the two local squeezing parameters are identical as $N_{1}=N_{2}$. It also presents that the performance can be enhanced with the local squeezing on the idler mode, which is related with the means and variances observed in the dHTD receiver. Under $N_{1}=N_{2}$, the performance of QI with the ASTM state is enhanced with the increasing squeezing parameters, as shown in Fig.~\ref{FigROC}(b).

\begin{figure}[t!]
	\centering	
	\includegraphics[width=0.35\textwidth]{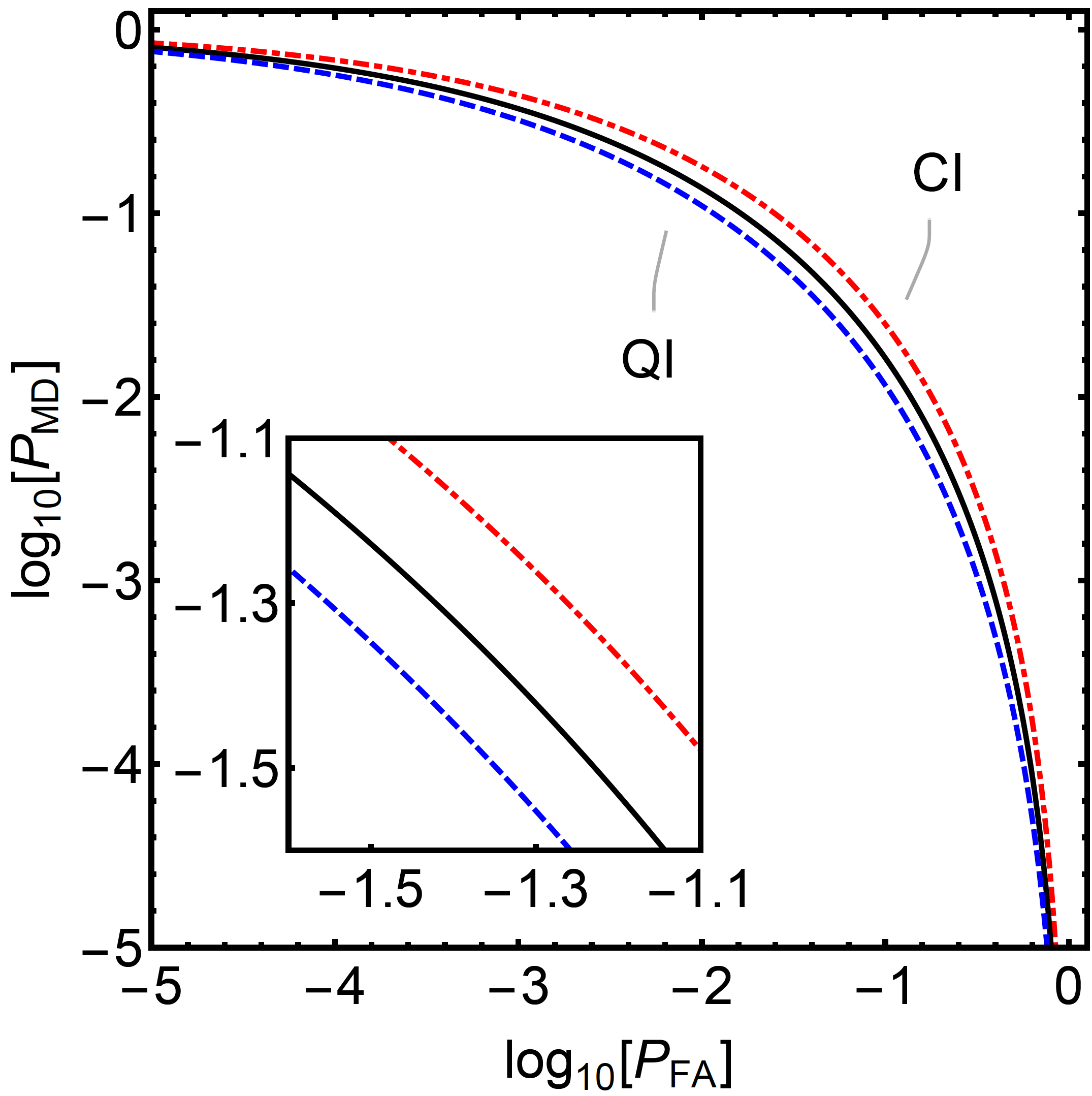}
	\caption{ROC curves of the conventional QI using a bright TMSV state with the dHTD receiver (blue dashed line), QI using an ASTM state with the dHTD receiver (black solid line), and CI with homodyne detection (red dot-dashed line). The parameters are the same as those of Fig.~\ref{FigAng}.}\label{FigROCQICI}
\end{figure}

Under a fixed mean photon number $N_{S}=2$ in the signal mode, in Fig.~\ref{FigROCQICI}, we compare three ROC curves: the conventional QI using a bright TMSV state, the CI using a coherent state with homodyne detection, and the QI using an ASTM state that is generated from the TMSV state with $N_{0}=1$. We obtain that the QI with the ASTM state outperforms the CI. However, it cannot beat the conventional QI with the bright TMSV states under a fixed $N_{S}$.

\section{Summary and discussion}\label{SecCon}

The aim of this work is to propose a method for improving the performance of QI under current technologies. Although the quantum advantage of QI improves under low signal energy, it is necessary to increase a signal energy for a remote target detection. However, the generation of a bright TMSV state is challenging, so that we need other ways to increase the energy of the signal.

With the optimal receiver, we found that a local squeezing on the signal mode which interacts with a target can improve the performance of QI, whereas a squeezing on the idler mode has no effect. On the other hand, with the dHTD receiver, a local squeezing on the idler mode cannot be neglected, and the symmetric local squeezing operation on both modes is the best strategy. We also observed that there is a threshold of the mean photon number of the initial TMSV state in order to beat the CI in QI using the ASTM state. Finally, we showed that our protocol provides a counterexample against the previous claim that quantum advantage of Gaussian QI is based on the quantum discord \cite{Bradshaw2017}.

One may wonder whether local squeezing operations on the return mode also can improve the performance of QI. In the QCB analysis, as it was described, unitary operations just before entering a receiver would be included in the optimal receiver so that there is no change in the performance under the local squeezing operation on the return mode. With the dHTD receiver, we could not find any improvement either, since the local squeezing on the return mode amplifies the transmitted thermal noise more than the reflected signal.

\begin{acknowledgements}
This work was supported by a grant to Defense-Specialized Project funded by Defense Acquisition Program Administration and Agency for Defense Development.
\end{acknowledgements}

\bibliographystyle{plain}

\onecolumn\newpage
\appendix

\section{Mean and variance of QI using ASTM state with dHTD receiver}\label{SecApp1}
Since we consider many ensembles for the target detection, the measurement results with the dHTD approach Gaussian distributions. The mean and variance can be calculated from Eq.~(\ref{state}) and Eq.~(\ref{dHet}), and they are written in the following equations:
\begin{align}
R&=\frac{1}{2} (\gamma_{1+}\gamma_{2+}+\gamma_{1-}\gamma_{2-}) C \sqrt{\kappa},\\
\Delta R&=\frac{1}{4}\left[2+A B(1-\kappa)(\gamma_{2+}^{2}+\gamma_{2-}^{2})+\kappa (A^{2}+C^{2})(\gamma_{1+}^{2}\gamma_{2+}^{2}+\gamma_{1-}^{2}\gamma_{2-}^{2})\right],
\end{align}
where $A=2N_{0}+1$, $B=2 N_{B}/(1-\kappa)+1$, $C=2\sqrt{N_{0}(N_{0}+1)}$, $\gamma_{j\pm}=\sqrt{N_{j}+1}\pm\sqrt{N_{j}}$, and $\kappa$ is the target reflectivity.

\end{document}